\documentclass[prl,twocolumn,amsmath,amssymb]{revtex4}

\usepackage{graphicx}
\usepackage{subfigure}
\usepackage{dcolumn}
\usepackage{array}
\usepackage{bm}
\usepackage[hang]{footmisc}
\def\ub{\underline}

\begin{document}

\title{\bf{Hexagonal $ABC$ semiconductors as ferroelectrics} \\[11pt] }
\author{Joseph W. Bennett, Kevin F. Garrity, Karin M. Rabe and David Vanderbilt}
\affiliation{Department of Physics and Astronomy\\
Rutgers University, Piscataway, NJ 08854 }

\date{\today}

\maketitle

{\bf We use a first-principles rational-design approach to identify a
previously-unrecognized class of ferroelectric materials in the
$P6_{3}mc$ LiGaGe structure type. We calculate structural parameters,
polarization and ferroelectric well depths both for reported and
as-yet hypothetical representatives of this class. Our results provide
guidance for the experimental realization and further investigation of
high-performance materials suitable for practical applications.}

A rapidly developing paradigm for the rational design of functional
materials is based on the first-principles study of large families of
known and as yet unreported compounds. First-principles calculations
of structure and properties are used first to explore the microscopic
origins and establish design principles for the functional properties
of interest, and then to screen a large number of both equilibrium and
metastable phases to identify promising candidate
systems~\cite{Jain11p2295, Armiento11p014103, Zhang12p1425,
Roy11preprint}. One recent study showed the semiconducting members of
the $ABC$ half-Heusler family to be piezoelectric, with a range of
piezoelectric properties comparable to the much-studied $AB$O$_{3}$
perovskite oxides~\cite{Roy11preprint}.

A ferroelectric is a material with a polar phase produced by a
structural transition from a nonpolar high-symmetry paraelectric
state, with an electric polarization that can be switched between two
or more symmetry-related variants by application of an electric
field~\cite{Lines77}. The rational design of new ferroelectric
materials is motivated both by fundamental scientific interest and by
potential technological applications~\cite{Izyumskaya07p111}. New
materials can offer better performance, including reduction in
switching time, in coercive field and in fatigue, operation at higher
or lower temperatures, and the possibility of better integration with
other materials based on structural or chemical compatibility.  New
ferroelectrics with lower band gaps for photoactive
applications~\cite{Bennett08p17409, Bennett10p184106, Gou11p205115}
are also of interest. Additional practical advantages could include
decreased toxicity, for example Pb-free~\cite{Bennett11p144122}, and
possible multifunctionality.

\begin{figure}
     \centering
     \subfigure[$P6_{3}/mmc$]{
          \label{fig:P63mmc}
          \includegraphics[width=1.50in]{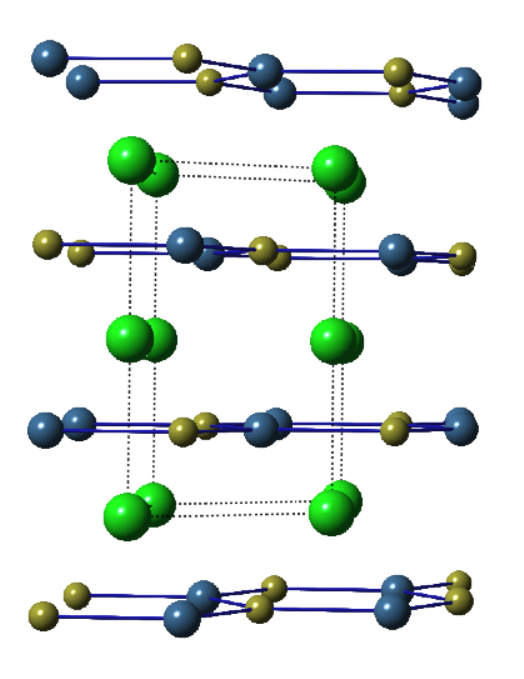}}
     \subfigure[$P6_{3}mc$]{
          \label{fig:P63mc}
          \includegraphics[width=1.50in]{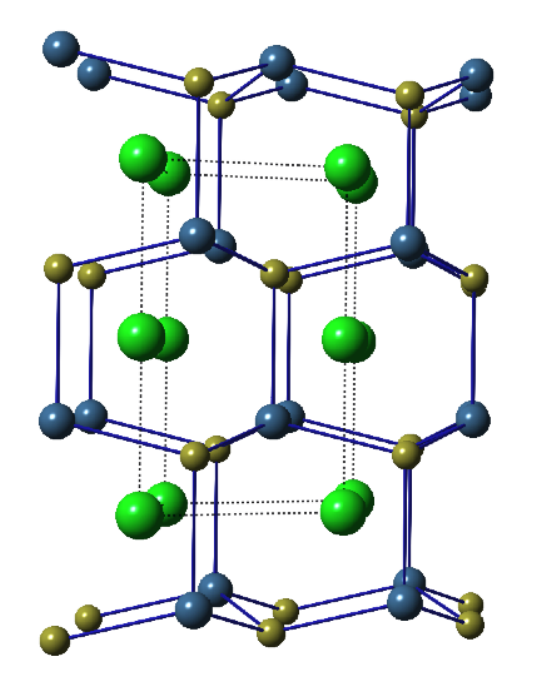}}
     \caption{The $P6_{3}/mmc$ ZrBeSi structure shown in a) is the
     nonpolar high-symmetry reference structure for the polar
     $P6_{3}mc$ LiGaGe structure shown in b). The two structures are
     related by a buckling of the planes formed by atoms at Wyckoff
     positions $2b$ (dark blue) and $2b^{\prime}$ (gold) and
     displacements of the planes relative to the stuffing atom at $2a$
     (green).}
       \label{fig:displacements}
\end{figure}

Any polar structure (if insulating) could potentially support
ferroelectricity if the barrier to switching is low
enough~\cite{Abrahams88p585, Abrahams06p26, Bennett12preprint}. We
therefore can search for new ferroelectric semiconductors by targeting
intermetallic compounds in polar space groups and screening both
reported and hypothetical compounds to find insulating representatives
with a low barrier to uniform switching through a nonpolar reference
phase, which provides an indication of the barrier to realistic
switching.  $ABC$ compounds with polar space group $P6_{3}mc$ in the
LiGaGe structure type~\cite{Bockelmann74p233, Burdett90p12,
Bojin03p1653, Casper08p035002} are a promising target class.  This
structure, shown in Figure~\ref{fig:displacements}, is a hexagonal
variant of the half-Heusler structure and can be described as a
wurtzite structure ``stuffed'' with a third
cation~\cite{Casper09p1090}.  The {\it Inorganic Crystal Structural
Database} (ICSD)~\cite{Belsky02p364} includes 18 $ABC$ compounds in
this structure type that do not contain an $f$-block element. We can
classify these combinations into the following groups: I-III-IV
(LiGaGe), I-II-V (LiBeSb), I-XII-V (LiZnSb), XI-III-IV (CuYSn),
XI-II-V (AgCaBi) and II-XII-IV (CaZnSn). In addition, we find that six
entries (CuScSn, CuYSn, AuYSn, AgCaBi, CaZnSn and CaHgSn) are also
reported with non-polar $P6_{3}/mmc$ symmetry in the ZrBeSi structure
type, which we identify as the nonpolar reference phase.  It has been
previously noted that the $P6_{3}/mmc$ ZrBeSi structure can be
obtained by a symmetry-restoring distortion of the LiGaGe structure in
which the buckling of the atomic planes in the wurtzite structure is
eliminated; this relationship is analogous to that of the wurtzite
structure to the metastable hexagonal structure of ScN
\cite{Farrer02p201203, Ranjan03p257602}.

In this paper, we use first-principles methods to establish a new
class of ferroelectrics in the LiGaGe structure type and to identify
promising candidate materials for further investigation.
Specifically, we compute the structural parameters, band gap,
polarization, and barrier to uniform switching of the eighteen
reported and 70 as-yet-hypothetical $ABC$ compounds in the LiGaGe
structure type. We identify several insulating combinations with
polarization comparable to or greater than that of BaTiO$_{3}$, and
uniform switching barriers comparable to or less than that of
PbTiO$_3$.  For all insulating combinations studied, we find that the
band gaps are in the semiconducting range; the lower band gaps could
be useful for photoactive applications~\cite{Bennett08p17409,
Bennett10p184106, Gou11p205115}. These candidate ferroelectrics offer
promise for experimental investigation and for the future development
of new high-performance materials for practical applications.

First principles computations were performed with the ABINIT
package~\cite{Gonze09p2582}. The local density approximation (LDA) and
a $4\times4\times4$ Monkhorst-Pack sampling of the Brillouin
zone~\cite{Monkhorst76p5188} were used for all calculations, except
for the Berry phase polarization~\cite{KingSmith93p1651, Resta94p899}
calculations, for which a $8\times8\times8$ grid was used. All atoms
were represented by norm-conserving optimized~\cite{Rappe90p1227}
designed nonlocal~\cite{Ramer99p12471} pseudopotentials, generated
with the OPIUM code~\cite{Opium}. All calculations were performed with
a plane wave cutoff of 50~Ry.

We first consider the eighteen non-rare-earth compounds that have
been experimentally reported in the LiGaGe structure type in the ICSD
\footnote{KSnAs is also reported in the LiGaGe structure type, but on
closer examination we find it to be misassigned. This will be
discussed in detail in a future publication.}.  For each combination,
we optimized the structural parameters for each of the three
structural variants $\ub{A}BC$, $A\ub{B}C$, and $AB\ub{C}$, where the
underscore indicates the element at Wyckoff position $2a$, which
stuffs the wurtzite structure comprised by the other two elements.
The results for the lowest energy structural variant are reported in
Table~\ref{table:exptdata}.  The computed structural parameters
generally show good agreement with experimental values, with the
underestimate of lattice constants characteristic of LDA calculations,
about 1-2~$\%$ for $a$ and as large as 3-4~$\%$ for $c$.

There are only two insulating compounds in the set of reported
LiGaGe-type compounds: LiBeSb (I-II-V) and LiZnSb (I-XII-V); in each
the stuffing atom is the monovalent element Li.  The total of 8 $s$
and $p$ valence electrons is expected to improve the likelihood of
band gap formation~\cite{Kandpal06p776}. The band gap for LiBeSb is
1.71 eV (indirect) and for LiZnSb it is 0.67 eV (direct). The computed
polarizations of 0.59 C/m$^2$ (LiBeSb) and 0.56 C/m$^2$ (LiZnSb) are
larger than that of BaTiO$_{3}$. To assess switchability, we compute
the energy difference between the polar state and the nonpolar
high-symmetry reference state.  While we recognize that ferroelectrics
do not switch by uniform change of the polarization through the
high-symmetry state, the energy barrier for uniform switching can be
used to assess the possibility of realistic switching by comparing
with the values for known ferroelectrics: 0.2~eV for PbTiO$_{3}$ and
0.02~eV for BaTiO$_{3}$ ~\cite{Cohen92p136}. In the present case, this
comparison suggests that the nominal barriers in LiBeSb and LiZnSb
(0.58 eV and 0.80 eV) are too high for switchability to be likely.

To search for candidate LiGaGe-type ferroelectrics with lower
barriers, we consider equiatomic combinations of three distinct
constituent elements $ABC$ with valences given by I-II-V or I-XII-V,
with I=(Li, Na, K), II=(Be, Mg, Ca, Sr, Ba), XII=(Zn) and V=(P, As,
Sb, Bi).  This generates a total of 72 candidate combinations to be
searched, only two of which are included in Table I.  We optimize the
structural parameters for each of the three variants corresponding to
the three choices of element for the $2a$ position.

We find that 6 of the 72 combinations are found to be metallic in the
lowest-energy structural variant: the computed structural parameters
and $\Delta E$, the energy relative to the relaxed high-symmetry
$P6_{3}/mmc$ phase, for these are given in
Table~\ref{table:hexdata1}. For the remaining 66 of the 72
combinations, the predicted lowest-energy variant is insulating, with
computed band gap ranging from 0.04 eV to 1.81 eV; since DFT tends to
underestimate band gaps, we expect that the actual fraction of
insulating compounds will be slightly higher than our calculations
would indicate. Of these 66 compounds, 49 have relaxed to the higher
non-polar $P6_{3}/mmc$ symmetry; results for these compounds are given
in the supplementary material. For the 17 polar insulating compounds,
we also compute the spontaneous polarization; results for these
compounds are given in Table~\ref{table:hexdata2}.

\begin{figure}
 \centering
\includegraphics[width=3.5in]{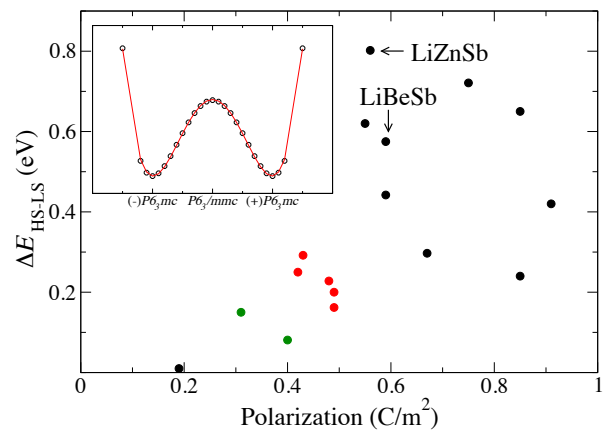}
\caption{Difference in energy between the low ($P6_{3}mc$) and high
($P6_{3}/mmc$) symmetry $ABC$ structures vs polarization for all polar
insulating combinations. Combinations with $A$=Li (black circles) are
less likely to be switchable than those with $A$=Na (red circles) or
$A$=K (green circles). The two reported compounds LiBeSb and LiZnSb
are labeled. The inset shows the characteristic ferroelectric double
well energy of NaMgP as a function of polar distortion obtained by
linear interpolation between the polar $P6_{3}mc$ and nonpolar
$P6_{3}/mmc$ structures.}
\label{fig:fig2}
\end{figure}

Thus, we have narrowed the search for new LiGaGe-type ferroelectrics
to seventeen polar insulating combinations.  For this set, we see in
Figure~\ref{fig:fig2} that $\Delta E$ has a positive correlation with
polarization, as would be expected in a simple double-well model. The
eight compounds LiBeP, LiCaBi, NaMgP, NaMgAs, NaZnSb, NaMgBi, KMgSb,
and KMgBi have polarizations comparable to that of BaTiO$_{3}$ and
$\Delta E < $0.25 eV, in the range favorable for ferroelectric
switching, and therefore are promising candidates for
ferroelectricity.

The ferroelectric double well for NaMgP, shown in the inset of
Figure~\ref{fig:fig2}, is representative of this group. The key to the
switchability of the polarization is that the wurtzite substructure is
not characterized by ideal $sp^3$ bonding like in ZnO, which would
require breaking and reforming of rigid bonds to switch. Rather, the
structure should be understood as the buckling of the flat planes of
the $P6_{3}/mmc$ structure, with $sp^2$ bonding. We define a buckling
parameter, $d$, as the distance along the $c$-axis between the
inequivalent atoms in the buckled plane. We find that $d$ decreases as
the size of the stuffing ion increases from Li to Na to K, weakening
the interplanar bond so that the barrier to switching is reduced most
for compounds containing K. This structural trend directly affects the
polarization, which arises from a combination of the buckling and the
displacement of the planes relative to the stuffing cation. Since
changing the sense of the buckling does not involve breaking and
reforming of bonds, the barrier to switching can be low enough for
ferroelectricity.

A key question is that of the prospects for synthesis of the candidate
compounds in the desired structure.  Of the eight compounds we have
identified as candidate ferroelectrics, six have reported structures
in ICSD. Five are reported in space group $P4/nmm$ (LiBeP, NaMgAs,
NaZnSb, KMgSb, KMgBi), and one in space group $Pnma$ (LiCaBi). Of the
two for which there is no reported structure, results from a recent
theoretical study~\cite{Zhang12p1425} predict them to be of $P4/nmm$
(NaMgP) or $P2_{1}/c$ (NaMgBi) symmetry.\footnote{Of the remaining
nine, two are reported with $P63mc$ symmetry (LiBeSb, LiZnSb), two in
space group $P4/nmm$ (LiBeAs, NaMgSb), four in $F{\bar4}3m$ (LiMgP,
LiZnP, LiMgAs, LiZnAs) and one predicted to be unstable (LiBeBi).}
However, it could still be possible to synthesize at least some of our
candidate LiGaGe-type ferroelectrics as metastable phases, in cases in
which the LiGaGe structure type is sufficiently close in energy to the
ground state. In particular, for NaZnSb the LiGaGe phase is only
0.04~eV per f.u. higher in energy than the $P4/nmm$ ground state,
which makes the metastable phase quite accessible. Furthermore, for
NaZnSb the energy difference between the lowest energy variant and the
next (with Zn as the stuffing atom) is 0.51~eV, suggesting that it
will be possible to obtain full chemical ordering.

If these compounds are grown as epitaxial films, this would provide an
additional route to engineering the polarization, the switching
barrier, and the relative stability of the LiGaGe phase. For example,
first-principles calculations show that 3$\%$ tensile strain in the
(0001) plane reduces the $\Delta E$ of the reported compounds LiZnP by
0.11~eV, LiZnAs by 0.12~eV and LiMgAs by 0.08~eV, to 0.57 eV, 0.60 eV
and 0.36 eV, respectively. Strain could also promote a polar
instability in the insulating nonpolar $P6_{3}/mmc$ compounds. Of the
forty-nine compounds we have identified as nonpolar insulators, six
are reported in ICSD with $P6_{3}/mmc$ symmetry (see supplemental
Table 1), and the previously mentioned theoretical study found five
additional compounds with this structure. First principles
calculations of the zone-center phonon frequencies for six selected
compounds (LiBaSb, NaBeSb, NaCaBi, KZnAs, KZnSb, and KBaSb) show that
in each case the frequency of the lowest frequency polar mode is below
100~cm$^{-1}$. However, the coupling of this mode to (0001) epitaxial
strain is not strong enough to produce an instability in the range
$\pm$~4~$\%$ in any of the six compounds we tested.

In conclusion, we have used first-principles methods to establish a
new class of ferroelectrics in the LiGaGe structure type and to
identify promising candidate materials for further investigation.
Through targeted synthesis, LiGaGe-type compounds could potentially be
developed as a valuable class of ferroelectric and piezoelectric
materials; other structure types with substructures related to
wurtzite could similarly yield systems with switchable polarization.
This is a specific application of a larger-scale strategy to identify
new ferroelectrics by targeting polar insulating compounds not
previously recognized as ferroelectric and tuning the composition and
other control parameters, such as epitaxial strain, and/or modifying
the structure by intercalation of atoms to reduce the barrier to
polarization switching.  The identification of ferroelectricity in
classes of materials in which it was previously unrecognized offers
the possibility of optimizing properties and combining polarization
with other functional properties, including magnetism, to produce
multifunctional behavior of fundamental scientific interest and for
groundbreaking technological applications.

\vspace{0.3cm}
\noindent{\bf Acknowledgments}
\vspace{0.3cm}

This work was supported in part by ONR Grants N00014-09-1-0302 and
N00014-05-1-0054. Calculations were carried out at the Center for
Piezoelectrics by Design.  We thank D. R. Hamann and R. Seshadri for
useful discussions. K. M. R. thanks R. Seshadri for hospitality at
UCSB and the Aspen Center for Physics (NSF Grant 1066293) where part
of this work was carried out.

\vspace{0.3cm}
\noindent{\bf Competing Financial Interests}
\vspace{0.3cm}

There are no competing financial interests.  


\begin{table}
\begin{center}
\begin{ruledtabular}
\begin{tabular}{lllllll}
& $a$ & $c$ & $z_{2b}$ & $z_{2b}^{\prime}$ & $\Delta E$\\
&(\AA)&(\AA)&          &                   & (eV)\\
\ub{Li}GaGe & 4.14 (4.18) & 6.71 (6.78) & 0.31 (0.31) & 0.70 (0.70) & 0.80\\
\ub{Sc}CuSn & 4.35 (4.39) & 6.80 (6.83) & 0.33 (0.33) & 0.73 (0.73) & 0.45 \\
\ub{Y}CuSn  & 4.48 (4.54) & 7.15 (7.27) & 0.32 (0.32) & 0.73 (0.73) & 0.32\\
\ub{Y}CuPb  & 4.51 (4.56) & 7.19 (7.33) & 0.32 (0.32) & 0.73 (0.73) & 0.34\\
\ub{Y}AgSn  & 4.63 (4.68) & 7.24 (7.37) & 0.31 (0.31) & 0.70 (0.72) & 0.39\\
\ub{Sc}AuGe & 4.26 (4.31) & 6.68 (6.85) & 0.30 (0.30) & 0.71 (0.70) & 0.61\\
\ub{Sc}AuSn & 4.51 (4.59) & 7.07 (7.20) & 0.35 (0.34) & 0.73 (0.73) & 0.75\\
\ub{Y}AuSi  & 4.25 (4.29) & 7.36 (7.55) & 0.27 (0.27) & 0.72 (0.73) & 0.05\\
\ub{Y}AuGe  & 4.36 (4.41) & 7.08 (7.31) & 0.28 (0.28) & 0.71 (0.72) & 0.26\\
\ub{Y}AuSn  & 4.61 (4.64) & 7.29 (7.37) & 0.33 (0.32) & 0.73 (0.73) & 0.70\\
\ub{Li}BeSb & 4.09 (4.15) & 6.64 (6.74) & 0.35 (0.34) & 0.73 (0.73) & 0.58\\
\ub{Li}ZnSb & 4.38 (4.43) & 7.08 (7.16) & 0.29 (0.30) & 0.67 (0.69) & 0.80\\
\ub{Li}ZnBi & 4.46 (4.58) & 7.23 (7.38) & 0.28 (0.28) & 0.66 (0.66) & 0.62\\
\ub{Ca}AgBi & 4.73 (4.81) & 7.56 (7.83) & 0.32 (0.31) & 0.72 (0.72) & 0.36\\
\ub{Ca}ZnSn & 4.60 (4.66) & 7.33 (7.63) & 0.31 (0.30) & 0.72 (0.72) & 0.32\\
\ub{Ca}HgSn & 4.75 (4.80) & 7.55 (7.76) & 0.31 (0.30) & 0.71 (0.72) & 0.53\\
\ub{Sr}HgSn & 4.88 (4.89) & 7.80 (8.22) & 0.29 (0.30) & 0.69 (0.72) & 0.36\\
\ub{Sr}HgPb & 4.89 (5.00) & 8.05 (8.17) & 0.29 (0.30) & 0.69 (0.72) & 0.19\\
\end{tabular}
\end{ruledtabular}
\caption{First-principles results for the eighteen LiGaGe-structure
compounds reported in ICSD. The computed ground state structural
parameters are compared to experimental values, given in parentheses
(in some cases, a transformation has been applied to facilitate
comparison). The origin is chosen so that the atom at position $2a$
(underlined) is at $z$=0. $\Delta E$ is the energy difference between
the polar state and the relaxed nonpolar high-symmetry reference
state.}
\label{table:exptdata}
\end{center}
\end{table}

\begin{table}
\begin{ruledtabular}
\begin{tabular}{lcccccc}
$ABC$&$a$&$c$&z$_{\rm 2b}$&z$_{\rm 2b^{\prime}}$&$d$&$\Delta E$\\
     &(\AA)&(\AA)&        &                     &   &(eV)   \\
Li\ub{Mg}Sb & 4.501 & 6.979 & 0.301 & 0.715 & 0.09 & 0.10 \\
Li\ub{Mg}Bi & 4.602 & 7.051 & 0.300 & 0.718 & 0.08 & 0.42 \\
\ub{Li}ZnBi & 4.461 & 7.233 & 0.277 & 0.658 & 0.12 & 0.62 \\
\ub{Na}ZnBi & 4.653 & 7.479 & 0.283 & 0.675 & 0.11 & 0.13 \\
\ub{K}SrBi & 5.757 & 7.044 & 0.250 & 0.750 & 0 & 0 \\
\ub{K}ZnBi & 4.582 & 10.135 & 0.250 & 0.750 & 0 & 0 \\
\end{tabular}
\end{ruledtabular}
\caption{First-principles results for the lowest-energy variant of
each of the 6 metallic $ABC$ combinations in the LiGaGe structure
type. The buckling parameter $d$ is described in the text. $\Delta E$,
as in Table 1.}
\label{table:hexdata1}
\end{table}

\begin{table}
\begin{ruledtabular}
\begin{tabular}{lcccccccc}
$ABC$&$a$&$c$&z$_{\rm 2b}$&z$_{\rm 2b^{\prime}}$&$d$&$E_{\rm gap}$&$\Delta E$&$P$\\
     &(\AA)&(\AA)&        &                     &   &(eV)         &(eV)      &(C/m$^2$)\\
\ub{Li}BeP & 3.634 & 5.833 & 0.295 & 0.686 & 0.11 & 1.51(i) & 0.24 & 0.85\\
\ub{Li}MgP & 4.134 & 7.211 & 0.413 & 0.773 & 0.14 & 1.81(i) & 0.42 & 0.91\\
\ub{Li}ZnP & 3.924 & 6.365 & 0.341 & 0.722 & 0.12 & 1.33(i) & 0.65 & 0.84\\
\ub{Li}BeAs & 4.091 & 6.636 & 0.265 & 0.647 & 0.12 & 1.71(i) & 0.30 & 0.67\\
\ub{Li}MgAs & 4.292 & 7.483 & 0.414 & 0.774 & 0.14 & 1.48(d) & 0.44 & 0.59\\
\ub{Li}ZnAs & 4.108 & 6.673 & 0.335 & 0.715 & 0.12 & 0.97(d) & 0.72 & 0.75\\
\ub{Li}BeSb & 4.094 & 6.645 & 0.267 & 0.650 & 0.12 & 1.13(i) & 0.58 & 0.59\\
\ub{Li}ZnSb & 4.376 & 7.081 & 0.288 & 0.669 & 0.12 & 0.67(d) & 0.80 & 0.56\\
\ub{Li}BeBi & 4.179 & 6.806 & 0.262 & 0.643 & 0.12 & 0.45(i) & 0.62 & 0.55\\
Li\ub{Ca}Bi & 4.679 & 7.503 & 0.287 & 0.731 & 0.06 & 0.16(i) & 0.01 & 0.19\\
\ub{Na}MgP & 4.424 & 6.877 & 0.310 & 0.716 & 0.09 & 1.17(d) & 0.20 & 0.49\\
\ub{Na}MgAs & 4.549 & 7.262 & 0.314 & 0.715 & 0.10 & 0.67(d) & 0.23 & 0.48\\
\ub{Na}MgSb & 4.868 & 7.584 & 0.307 & 0.705 & 0.10 & 0.69(d) & 0.29 & 0.43\\
\ub{Na}ZnSb & 4.558 & 7.332 & 0.285 & 0.677 & 0.11 & 0.20(d) & 0.16 & 0.49\\
\ub{Na}MgBi & 4.957 & 7.584 & 0.302 & 0.704 & 0.10 & 0.14(i) & 0.25 & 0.42\\
\ub{K}MgSb & 5.017 & 7.789 & 0.290 & 0.697 & 0.09 & 0.59(d) & 0.08 & 0.40\\
\ub{K}MgBi & 5.092 & 8.005 & 0.291 & 0.698 & 0.09 & 0.15(i) & 0.15 & 0.31\\
\end{tabular}
\end{ruledtabular}
\caption{First-principles results for the lowest-energy variant of
each of the 17 polar insulating $ABC$ combinations in the LiGaGe
structure type. The structure is specified by the lattice constants
and the internal structural parameters for two of the atoms; the
origin is chosen so that the atom at position $2a$ (underlined) is at
$z$=0.  The buckling parameter $d$ is described in the text.  The
computed values for the band gap $E_{\rm gap}$ (d=direct and
i=indirect), $\Delta E$ (as in Tables I and II), and polarization $P$
are included.}
\label{table:hexdata2}
\end{table}


\end{document}